# CIPM: Common Identification Process Model for Database Forensics Field

Ibrahim Alfadli
College of Computer Science and Engineering. Information system department. Taibah University. Saudi Arabia. Madina
ialfadli@taibahu.edu.sa

Fahad M Ghabban
College of Computer Science and Engineering. Information system department. Taibah University. Saudi Arabia. Madina
fghaban@taibahu.edu.sa

Omair Ameerbakhsh
College of Computer Science and Engineering. Information system department. Taibah University. Saudi Arabia. Madina
oameerbakhsh@taibahu.edu.sa

Amer Nizar AbuAli
College of Computer Science and Engineering. Information system department. Taibah University. Saudi Arabia. Madina
aabuali@taibahu.edu.sa

Arafat Al-Dhaqm
Faculty of Engineering, School of Computing,
University Technology Malaysia
Malaysia, Johor
mrarafat1@utm.my

Mahmoud Ahmad Al-Khasawneh
Faculty of Computer & Information Technology
Al-Madinah International University
Shah Alam, Malaysia
mahmoud@outlook.my

*Abstract*— Database Forensics (DBF) domain is a branch of digital forensics, concerned with the identification, collection, reconstruction, analysis, and documentation of database crimes. Different researchers have introduced several identification models to handle database crimes. Majority of proposed models are not specific and are redundant, which makes these models a problem because of the multidimensional nature and high diversity of database systems. Accordingly, using the metamodeling approach, the current study is aimed at proposing a unified identification model applicable to the database forensic field. The model integrates and harmonizes all exiting identification processes into a single abstract model, called Common Identification Process Model (CIPM). The model comprises six phases: 1) notifying an incident, 2) responding to the incident, 3) identification of the incident source, 4) verification of the incident, 5) isolation of the database server and 6) provision of an investigation environment. CIMP was found capable of helping the practitioners and newcomers to the forensics domain to control database crimes.

*Keywords— Database forensics; Identification process; digital forensics, metamodelling*

## I. INTRODUCTION

Database Forensics (DBF) is a field classified under the digital forensics field of study, which inspects database content with the aim of verifying database crimes [1][2]. DBF is well recognized as a field of high significance because it can identify, detecting, acquiring, analyzing, and reconstructing incidents that occur in a database, thereby determining the intruders' activities. On the other hand, it suffers from several problems due to which DBF is known as an unstructured field of heterogeneity and confusion [3]–[8][9]. For instance, it involves various database system infrastructures, the database systems are generally multidimensional, and the knowledge domain is scattered in all directions [10]. Different infrastructures of the database system with a multidimensional nature help DBF to deal with specific incidents. As a result, each database management system (DBMS) has different identification investigation model/approach. For that reason, the issues of forensic investigation processes[11] and the scattering of knowledge in all directions have led to some other problematic areas for practitioners and researchers working in the DBF field [3], [12], [13]. This knowledge includes models, techniques, processes, tools, activities, methods, frameworks, algorithms, and approaches) is generally disorganized and non-structured. In addition, this is dispersed at a universal level, for instance, on the Internet, journals, books, reports, online databases, conferences, dissertations, and organizations. As a result, the literature lacks identification models capable of unifying the concepts and terminologies in a way to lessen the existing confusion and help in making domain knowledge well organized and structured [14]–[20]. As various infrastructures exist for DBMS, several DBF process models have been introduced in the literature to manage the field of DBF from the perspective of the investigation process [21]–[24] Nonetheless, among them, no single model can be found to be applied as a common model to the DBF field. The process models are typically applied to solving certain database incidents, case studies, or scenarios. For that reason, these models have generated processes of high redundancy and irrelevancy. The models introduced in this field have addressed DBF from four perspectives of investigation processes: 1) identification process, 2) collection process, 3) analysis process and 4) presentation process. In the current study, the identification process perspective is discussed in detail. The models discussing DBF from an identification process perspective typically hold different and redundant investigation processes, tasks, and activities. The *Suspension of Database Operation* process provided in the model proposed in [25], [26] works by isolating the database server from the users to capture database activities. However, the *Verification* and *System Description* processes provided in the model introduced in [27] [28] works by verifying the database incidents, isolating the database server, confirming the incident, documenting the system information like the name of the system, operating system, serial number, system function, and physical descriptions. Moreover, the *Identification* process suggested in the models proposed in [29], [30] indeed disconnects the database server from the network to capture volatile data. Similarly, the *Incident verification* and *Investigation preparation* processes proposed in [31] [32] are aimed at identifying and verifying the database incidents by a preliminary investigation. These models then provide forensic workstations and toolkits to give a response to the incidents; then, they disconnect the database server. Additionally, the *Database Connection Environment* introduced by the authors in [33] prepares the investigation environment, gains permission for having access to the database, and executes the commands needed. Moreover, the *Table Relationship Search and Join process* is mainly aimed at extracting the tablespaces within the database, choosing the target[34], selecting the tables storing the investigation data, and checking recurrently the other table field. In the *Data Acquirement with Seizure and Search Warrant* process, there is a need to secure the evidence location and extract the evidence related to a certain crime or incident [33], [35]. *Server Detection* is another process of interest, which detects the server that runs a database system. This process grasps the overall network environment in a firm. It acquires the network topology in the firm, which is applied to the identification and detection of the victim database server [36]–[38]. The researchers in [39] described the *Setup Evidence Collection Server* process that was employed to prepare the investigation environment for storing the incidents, whereas the *Identification* process introduced in [40] is used to identify the relevant MySQL database files (log files, text files, binary files) and utilities. Likewise, in [41], [42], the *Incident reporting and Examination Preparation* processes were introduced to capture database incidents utilizing the users' reports, system audit, or triggered events. To discover the database incidents, they cut off the network, configure the investigation circumstance, identify policies, prepare the appropriate tools, and inform the decisions made. Furthermore, the authors in [43] proposed the *Determining Database Dimension* and *Acquisition Method process* to identify the dimensions of a database that has been hacked/attacked. After that, the appropriate acquisition methods for that dimension are determined. In addition, the *Choose environment and Select implement method* process suggested in [44] choose the forensic context (clean or found environment) and choose an approach to transforming the forensic setting into the chosen forensic setting. The *Preliminary analysis* process introduced in [37], [45], [46] was implemented with the aim of forming an architectural visualization of DBMS with all components and their locations in the layered model of DBMS, identifying the files and folders in the layers below the storage engines' layer, preparing and utilizing the forensic tools and processes for the creation of an initial image and then the collection of metadata values of the identified target files, and recording the metadata of the target files [15], [47], [48][49]. Accordingly, the current study is mainly aimed at integrating all exiting identification process tasks and activities into an identification process model applicable to DBF with the use of the metamodeling approach, called CIPM. It comprises three phases: 1) Pre-identification, 2) During-identification, and 3) Post-identification. The three phases, in turn, comprise a number of steps and processes aiming to provide the techniques, tools, strategies, policies, and methods of investigation [48], which are implemented for the purpose of identifying, gathering, preserving, analyzing, and documenting database crime. CIPM has the capacity of structuring, unifying, facilitating, managing, sharing, and reusing the DFI field knowledge amongst the practitioners working in the forensics field. The metamodeling approach identifies the general concepts existing within each problem domain and examines the relationships among them. It makes a less complex, less interoperable, and less heterogeneous domain [50], [51]. As a result, the metamodels/models need to be defined meticulously and structured properly.

Therefore, the rest of this paper is arranged as follows: Section 2 discusses the methodology and Section 3 presents the initial results and discussion, and finally, Section 4 concludes the whole study.

II. METHODOLOGY

In designing CIPM, five steps of the metamodeling creation process were adapted as presented in [52] (see Fig. 1).

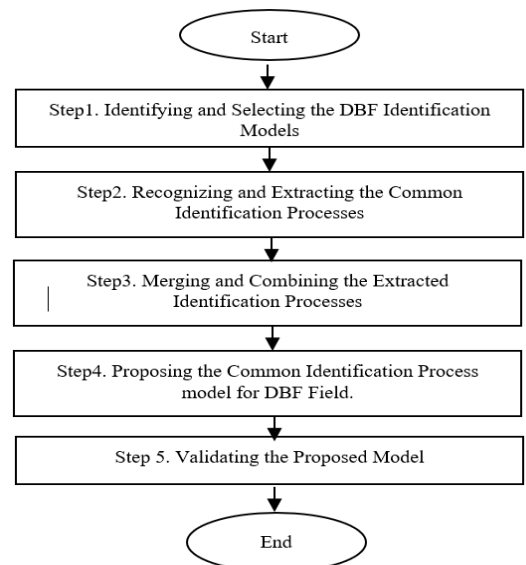

Fig. 1. Metamodeling development process

*Step 1. Identifying and selecting the DBF Identification Models:* This step identifies and selects the production and validation models. Several DBF identification models proposed in the current literature were analyzed [53]–[55] In this paper, the model selection process was done based on the coverage factors identified in former studies. This step results in 19 identification models that can be applied to developing and constructing purposes (see Table I).

TABLE I. DEVELOPMENT MODELS AND EXTRACTED PROCESSES

| Models | Database Forensic Identification Models | Similar Process |
|---|---|---|
| M1 | [25] | Suspend Database Operations |
| M2 | [27] | Verification |
| M3 | [29] | System Description |
| M4 | [31] | Identification process |
| M5 | [33] | Identification process |
| M6 | [56] | Investigation Preparation |
| M7 | [57] | Incident Verification |
| M8 | [36] | Database Connection Environment |
| M9 | [39] | Table Relationship Search and Join Process |
| M10 | [40] | Data Acquirement with Seizure and Search Warrant |
| M11 | [41] | Server Detection |
| M12 | [43] | Setup Evidence Collection Server |
| M13 | [58] | Incident reporting |
| M14 | [59] | Examination Preparation |
| M15 | [60] | Determine Database Dimension |
| M16 | [61] | Determining Acquisition Method |
| M17 | [62] | Identification |
| M18 | [63] | Preliminary analysis |

*Step 2. Recognizing and extracting the Common Identification Processes:* A DBF identification process resulted from the 18 models chosen earlier. Throughout this process, definite criteria were used for the identification of the relevant and appropriate identification processes. The criteria applied to determine the identification processes were taken from [64]–[66]. The criteria were used to avoid any missing or random process selections. Thus, in this section, a total of 18 identification processes were determined and derived from those 18 DBF identification models (see Table I).

*Step 3. Merging and Combining of the Extracted Identification Processes:* This step involves merging and combining the 18 extracted processes based on similarity in semantic meaning or functional meaning [67], [68]. All identification processes that have identical semantic meaning or functional meaning are organized into an abstract model called Common Identification Process Model (CIPM).

*Step 4.* Proposing CIPM for the DBF Field: This step introduces CIPM to be applied to the DBF field with the help of the mapping process. CIPM comprises three stages as demonstrated in Fig. 2.

*Step 5.* Validating the Proposed Model: CIPM is validated and enhanced to make it perfect and coherent as much as possible. To this end, the model is compared to identical existing models [67], [69].

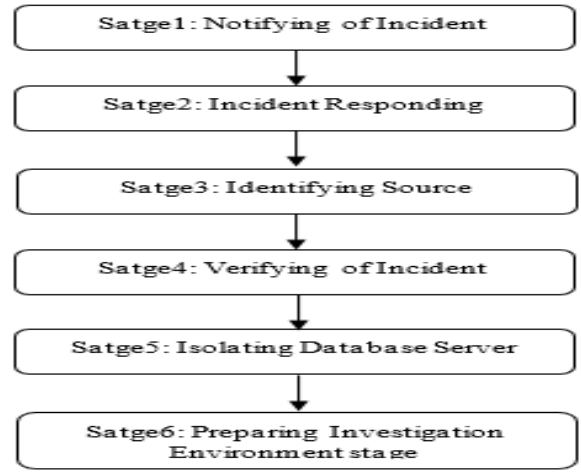

Fig 2. Common identification process model (CIPM)

III. RESULTS AND DISCUSSION

All DBF identification models, processes, tasks, and activities that appeared on the Internet, conferences, journals, online databases, books, and dissertations are specific and redundant. As a result, in the current paper, a common identification process model (CIPM) was designed in a way to be well applicable to the DBF field (see Fig. 1). It was attempted to include all the advantages of the previously proposed models in CIPM. It was properly defined in terms of meaning, tasks, and activities. This model is mainly aimed at solving the DBF field identification processes of irrelevancy and redundancy that tend to confuse practitioners working in the forensics domain. CIPM covers all the strengths of the models formerly proposed in the literature (see Table II). For example, the proposed Identification *process* contained whole investigation tasks, activities, and methods, of the eighteen (18) existing identification processes. Note that, in the present paper, *process* refers to respective stages that a scientific framework/model applies to the achievement of the most important goals, which is, in the case of the current study, producing an admissible forensic hypothesis applicable during the litigation process [54], [70].

TABLE II. A COMPARATIVE SUMMARY: PROPOSED COMMON IDENTIFICATION PROCESS AGAINST VALIDATION MODELS

| Proposed Common DBF Process | Existing DBF Models | | | | | | | | | | | | | | | | | |
|---|---|---|---|---|---|---|---|---|---|---|---|---|---|---|---|---|---|---|
| | M1 | M2 | M3 | M4 | M5 | M6 | M7 | M8 | M9 | M10 | M11 | M12 | M13 | M14 | M15 | M16 | M17 | M18 |
| Identification process | √ | √ | √ | × | √ | √ | × | √ | √ | √ | √ | × | √ | √ | × | × | × | × |

As can be seen in Fig. 1, this comprises six stages: notifying of the incident, responding to an incident, identification of the source, verification of incident, isolation of the database server, and preparation of investigation environment. In the first stage, the DBA of the firm notifies the higher management staff (for instance, Chief Executive Officer (CEO), Chief Security Officer (CSO), Chief Operating Officer (COO)) concerning the incident occurred to the database server [61]. When it occurs, the firm CEO has two choices [31]: assigning an internal/external investigation team for performing required investigations on the incident or stopping the investigation [31]. In case of the first option, the team will perform the second stage of the identification investigation process (i.e., the Incident Responding Stage) with the aim of collecting detailed data related to the incident, e.g., information in regard to the incidence of the event, parties involved, and the number of the databases involved associated with their sizes [27], [31]. To this end, the team applies forensic techniques for the purpose of seizing the investigation sources [33], collects the instable artefacts [29], and also collect helpful information by holding different interviews with the staff [36]. Incident Responding [30], [71], in turn, comprises three concepts: Capture, Live Response, and Interview. Accordingly, the investigation team captures the investigation sources, for instance volatile and non-volatile artefacts [29]. In addition, the Live Response is related to the concept of Volatile Artefact. As a result, the team collects useful volatile data from the Volatile Artefact concept. The last concept addressed in the Incident Responding is Interview. The team is required to carry out some interviews with the firm's senior staff (e.g., CEO and DBA) [36]. These interviews may result in collecting basic information, e.g., information accounts, database server, users, network ports, incident reports, investigation processes, logs, and policies [36]. It is clear that the incident responding stage helps the team determine an incident boundary and also decide about the investigation sources [72], [73]. The third stage involves Identifying Source Stage through which specific investigation sources are identified [29], [31], [40]. In an investigation source, there may be several helpful volatile and non-volatile artefacts that hold valuable evidence. As a result, this stage addresses the concepts captured during the responding stage (e.g., Artefact, Volatile Artefact, Nonvolatile Artefact, Source, Log File, Database File, and Undo Log concept). Then, the next investigative stage is dedicated to Verifying of Incident stage, which helps the team verify the incident occurred to the database [27], [31]. This stage comprises nine main concepts: Investigation Team, Incident, Destroyed Database, Modified Database, Compromised Database, Incident Types, Report, Company, and Decision. Thus, the team need to identify the type of the incident (is it compromised, destroyed, or modified) [43], the nature of the incident, and the status of the incident. After that, the team will submit a detailed report the firm management in regard to the incident to [33]. Managers will review the report submitted and, accordingly, make necessary decisions; three options exist: keeping on the investigation process, stopping the process, or disconnecting the database server from the network [29], [41]. When the incident is completely verified and determined, then fifth stage, i.e., Isolating Database Server, starts. Through this stage, the team has the ability of isolating/disconnecting [72], [73] a suspect database server from the network with the aim of avoiding more tampering [25], [31]. This stage comprises three concepts: Investigation Team, Database Server, and Database Management System. Note that isolating or disconnecting the suspect database server does not necessarily result in the database shutdown [29]; rather, it isolates users from the database management system [25], [41]. At the final stage, the team need to carry out the task of Preparing Investigation Environment [74]. It makes the team capable of preparing the investigation setting in which the whole investigation task could be carried out in an effective way [31]. The investigation setting involves six different concepts: Investigation Team [75], Forensic Workstation, Found Environment, Clean Environment, Forensic Technique, and Source. The team provides the trusted forensic workstation involving the trusted forensic technique (forensic tools and methods) and the investigation sources identified formerly through the identification stage [11].

IV. CONCLUSION

The present study identified 18 identification processes derived from eighteen (18) DBF models. Then, the extracted processes were collected and refined on the basis of common objectives with the aim of identifying a common identification process model applicable to the database forensics field. It ended with proposing the Common Identification Process Model (CIPM). The identified processes and models were comprehensively investigated; based on the results obtained, CIPM showed that six stages are commonly involved in the database forensic investigation process: notifying of incident, incident responding, identifying source, verifying of incident, isolating database server, and preparing the investigation setting. Afterward, the perfectness of CIPM was confirmed with the use of the existing database forensic investigation models. The studies conducted into this subject in future could 1) analyze the concepts and relationships that exist within each of the identified stages in CIPM through the use of a software engineering approach that is recognized as metamodel, and 2) evaluate CIPM by means of the real-world case studies.